\documentclass[11pt]{JHEP3}
\usepackage{graphicx}
\usepackage{amsmath}
\usepackage{braket}

\newcommand{\be}{\begin{equation}}
\newcommand{\ee}{\end{equation}}
\newcommand{\bea}{\begin{eqnarray}}

\newcommand{\eea}{\end{eqnarray}}
\newcommand{\ti}{\widetilde}

\newcommand{\gl}{\lambda}
\newcommand{\refe}[1]{Eqn.~(\ref{#1})}

\title{\begin{center} 
Hybrid Gauge Mediation 
\end{center}}
\author{
~~ Moritz McGarrie\\
Queen Mary University of London\\
Center for Research in String Theory\\
Department of Physics\\
Mile End Road, London, E1 4NS, UK.\\

\email{m.mcgarrie@qmul.ac.uk}
}

\abstract{Inspired by four dimensional (de)constructions, we use the framework of ``General gauge mediation in five dimensions'' to interpolate between gaugino and ordinary gauge mediation. In particular we emphasise that an intermediate hybrid regime of mediation may be obtained in these higher dimensional models as has been obtained in the quiver gauge models.}

\preprint{QMUL-PH-11-01}
\keywords{General gauge mediation, Gaugino mediation, Deconstruction}

\begin{document}
\section{Introduction}

\begin{flushleft}
\end{flushleft}
\begin{figure}[ht]
\centering
\includegraphics[scale=0.4]{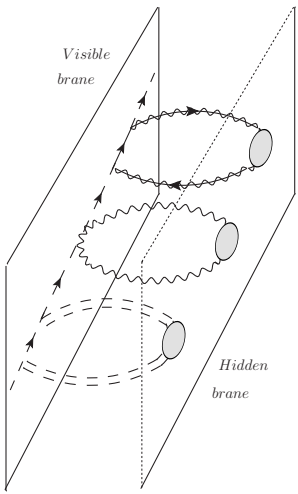}
\caption{A pictorial of the leading order ($\alpha^2$) sfermion mass contributions due to general gauge mediation across an interval.  Sfermion mass contributions on the visible brane are generated by supersymmetry effects encoded in current correlators (blobs) located on the hidden brane.}
\label{flatfigure}
\end{figure}
In gauge mediated supersymmetry breaking, one may broadly split models into two classes, those that are gauge mediated and those that are RG gaugino mediated.  To be more precise, and using the construction of general gauge mediation (GGM) \cite{Meade:2008wd},  the soft term for scalars masses at lowest order in the standard model coupling $\alpha$ is dependent on a super-traced set of current correlators 
\be 
[3\tilde{C}_1(p^2/M^2)-4\tilde{C}_{1/2}(p^2/M^2)+\tilde{C}_{0}(p^2/M^2)] 
\ee
as pictured in figure \ref{flatfigure}. In general, given a perturbative hidden sector, we may expand this set of current correlators in either of two expansions to obtain analytic expressions:  the four dimensional gauge mediated limit $M^2/p^2\leq 1$ or the screened (five dimensional) limit $p^2/M^2<1$ \cite{Mirabelli:1997aj,McGarrie:2010kh,McGarrie:2010qr,McGarrie:2010yk}. In this second limit the scalar soft masses are rather suppressed at the high characteristic scale $M$ of the hidden sector. However, due to the gaugino mass contributions in the renormalisation group equations (RGE's)\footnote{See \cite{Drees:2004jm,Martin:1993zk,Yamada:1994id} for the four dimensional RG equations.} \cite{Bhattacharyya:2010rm} as in figure 2, scalars develop soft masses at low scales through RG gaugino mediation \cite{Schmaltz:2000gy,Schmaltz:2000ei}.  We should of course point out the mechanism by which one discriminates between the two expansion limits: when a mass scale $m_v$ enters into the outer loop (see figure \ref{flatfigure}) of the leading order sfermion mass diagrams with $m_v \ll M$,  this mass scale suppresses loop momenta and warrants expanding the current correlators in the screened limit. For example, five dimensional models of super Yang-Mills on an $R^{1,3}\times S^1/\mathbb{Z}_{2}$ background with interval length $\ell=\pi R$ will have Kaulza-Klein (kk) masses $m_{n}=\frac{n\pi}{\ell}$. As higher dimensional models naturally introduce this mass scale through the kk modes,  it is customary that the screened limit become synonymous with higher dimensional mediation of supersymmetry breaking.

It is natural to ask if there is some intermediate type of mediation whereby the leading order scalar soft masses are still somewhat suppressed at the high scale but not as drastically as in the gaugino mediation limit, such that both the leading order contributions and subleading RG contributions play a significant role: hybrid gauge mediation.  The hybrid regime, when $m_{v}\sim M$, cannot be accessed by expanding the current correlators in $M^2/p^2$ and one must find a new way to evaluate the sfermion diagrams.  However for the case of a minimally truncated kk tower with just one massive kk mode, the leading orders sfermion diagrams are analytically solvable for all ratios of $F$, $M$ and $m_{v}$ and one can not only interpolate between the four dimensional and screened five dimensional limit, but one can also access the hybrid regime.
\begin{flushleft}
\end{flushleft}
\begin{figure}[htcb]
\centering
\includegraphics[scale=0.6]{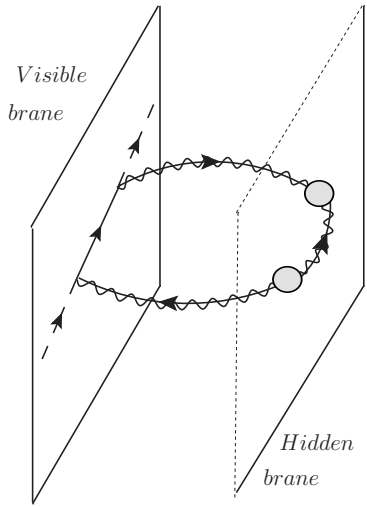}
\caption{A diagram representing a subleading (order $\alpha^3$) contribution to sfermion masses ($p_{ext}=0$) through a gaugino mediated double mass insertion of the Majorana soft mass on the hidden brane. For non-zero external scalar momenta, this diagram is a correction to the scalar kinetic term, leading to the soft mass contribution to one loop renormalisation group equations (RGE's) for sfermions.}
\label{figure2}
\end{figure}
This hybrid regime has been explored for a (de)construction model \cite{Auzzi:2010xc,Auzzi:2010mb}.  It also naturally arises in ISS-like models that exhibit colour-flavour locking that generate linking fields \cite{Green:2010ww,Komargodski:2010mc}. To some extent, phenomenological scans which implement hybrid gauge mediation have also already been explored \cite{Abel:2010vba,Abel:2009ve,Rajaraman:2009ga,Thalapillil:2010ek} by virtue of having a nonzero scalar soft mass at a high scale and effective four dimensional RG equations\footnote{The assumption of four dimensional RG equations is not always valid: it is dependent on the relative choice of energy scales as depicted in figure \ref{energyscales}.}.  Our interest in this paper is to demonstrate the analytic calculability of the visible brane localised scalar and bulk scalar soft masses in this five dimensional model. In this paper we would like to explore the hybrid regime in the five dimensional construction of general gauge mediation and also to document further progress pertaining to both the five dimensional and (de)construction model \cite{Auzzi:2010xc,Auzzi:2010mb}. 

\emph{The key message} of this paper is that whereas gaugino mediated models typically give a leading order sfermion mass and bulk scalar mass of 
\be
m^2_{\tilde{f}(5d)} \sim  m^2_{\tilde{f} (4d)}\frac{1}{(M\ell)^2},
\ee
more generally by changing the ratio of the first kk mass $m_{v}=\frac{\pi}{\ell}$ with $M$ one can obtain analytic formulas whereby
\be
m^2_{\tilde{f} (5d)} \sim m^2_{\tilde{f} (4d)} \frac{1}{(M\ell)^\rho}
\ee
where $\rho$ takes real values between $0$ and $2$.

This paper is organised as follows: Section \ref{section:Framework} outlines the framework of general gauge mediation in five dimensions.  Section \ref{section:truncated}  explores a truncated limit of this model using the language of current correlators and is quite general.  We make use of these results in section \ref{section:hybrid}   where we specify the hidden sector matter content to be a generalised messenger sector coupled to a spurion.  Specifying the hidden sector determines the structure of the current correlators of section \ref{section:truncated} which form part of two loop diagrams which are then analytically solved for all ratios of $F$, $M$ and $m_{v}$.  Section   \ref{section:subleading} comments on the contribution of a particular subleading diagram to the leading order masses and in section \ref{conclusion} we conclude.  The apppendix  \ref{appendixA} contains a more detailed computation of the main result of this paper.
\section{Framework}\label{section:Framework}
First we will recall the essential features of $\mathcal{N}=1$ super Yang-Mills in $5d$, compactified on  $R^{1,3}\times S^{1}\!/\mathbb{Z}_{2}$ \cite{Mirabelli:1997aj,Hebecker:2001ke,McGarrie:2010kh}.  The $\mathcal{N}=1$ action written in components is
\be
S_{5D}^{SYM}= \int d^{5} x
~\text{Tr}\left[-\frac{1}{2}(F_{MN})^2-(D_{M}\Sigma)^2-i\bar{\gl}_{i}\gamma^M
D_{M}\gl^{i}+(X^a)^2+g_{5}\, \bar{\gl}_i[\Sigma,\gl^i]\right].
\ee
The five dimensional coupling $1/g^2_{5}$ has been rescaled inside the covariant derivative, $D_{M}=
\partial_{M}+ig_{5} A_{M}$, which is useful for perturbative computations. The other fields are a real scalar $\Sigma$, an $SU(2)_{R}$ triplet of
real auxiliary fields $X^{a}$, $a=1,2,3$ and a symplectic Majorana
spinor $\gl_{i}$ with $i=1,2$ which form an $SU(2)_R$ doublet. The reality condition is $\gl^i= \epsilon^{ij} C\bar{\gl}_{j}^{T} $.
The interval $S^{1}\!/\mathbb{Z}_{2}$, boundaries at $y=0,\ell$ will
preserve only half of the $\mathcal{N}=2$ symmetries and we choose to
preserve the supersymmetry parameter $\epsilon_{L}$ and set $\epsilon_{R}=0$.  We have a preserved parity
whereby  $P\psi_{L}=+\psi_{L}$ $P\psi_{R}=-\psi_{R}$ for all fermionic fields
and susy parameters.  The action of the parity is such that $\mathbb{P}\Phi(y)=P\Phi(-y)$ for all the fields. This is equivalent to requiring Neumann and Dirichlet boundary conditions, respectively.  One can then group the
susy variations under the positive parity assignments and they fill an off-shell $4d$ vector
multiplet $V(x^{\mu},x_{5})$.  Similarly the susy variations of odd parity
form a chiral superfield $\Phi(x^{\mu},x_{5})$. The Kaluza-Klein masses $m_{n}$ arise from $p_{5}=\frac{n\pi}{\ell}$. We may therefore write a $5d$
$\mathcal{N}=1$ vector multiplet as a $4d$ vector and
chiral superfield:
\begin{alignat}{1}
V=&- \theta\sigma^{\mu}\bar{\theta}A_{\mu}+i\bar{\theta}^{2}\theta\gl-
i\theta^{2}\bar{\theta}\bar{\gl}+\frac{1}{2}\bar{\theta}^{2}\theta^{2}D\\
\Phi=& \frac{1}{\sqrt{2}}(\Sigma + i A_{5})+
\sqrt{2}\theta \chi + \theta^{2}F\,,
\label{fields}
\end{alignat}
where the identifications between $5d$ and $4d$ fields are
\begin{equation}
D=(X^{3}-D_{5}\Sigma) \quad F=(X^{1}+iX^{2})\,,
\end{equation}
and we used $\lambda$ and $\chi$ to indicate $\lambda_{L}$ and
$-i\sqrt{2}\lambda_R$ respectively. The bulk fields have a bulk propagator given by 
\begin{equation}
  \left\langle{ a(x,x^5) a(y,y^5)} \right\rangle = 
	\int_{p5} \frac{i }{p^2 - (p_5)^2}\,
 e^{-ip\cdot (x-y)} (e^{ip_5(x^5-y^5)}
     + P e^{ip_5(x^5+y^5)}) \ ,
\label{fiveprop}\end{equation}
where 
\begin{equation}
   \int_{p5} = \int \frac{d^4p}{ (2\pi)^4} \frac{1}{ 2\ell}\sum_{p_5}. \ 
\label{intdef}\end{equation}
The vector and chiral superfield have a Kaluza-Klein (kk) mode expansion given by 
\begin{alignat}{1}
V(x,y)= &\frac{1}{\sqrt{\ell}}V^0 (x)+\frac{\sqrt{2}}{\ell}\sum^{\infty}_{n=1}V^n (x) \cos \frac{n\pi y}{\ell}\\
\Phi(x,y)= &\frac{\sqrt{2}}{\ell}\sum^{\infty}_{n=1}\Phi^n (x) \sin \frac{n\pi y}{\ell}.
\end{alignat}
Next, we locate a supersymmetry breaking hidden sector on one fixed point of the interval (the hidden brane) and encode the effects in terms of a brane localised  $\mathcal{N}=1$ 4d current superfield.   We gauge the global symmetry associated with these currents and couple to the positive parity vector superfield in the bulk:
\begin{equation}
\mathcal{S}_{int}=2g_{5}\!\int\! d^5x d^{4}\theta \mathcal{J} 
\mathcal{V}\delta(x_{5})= \!\int\! d^5x g_{5}(JD- \gl j \!-
 \!\bar{\gl} \bar{j}-j^{\mu}A_{\mu})\delta(x_{5}).
\end{equation}
For the leading order soft masses,  due to the fixed points breaking translation invariance, one obtains an infinite kk tower of Majorana masses that couple each kk mode to every other given by 
\be 
\mathcal{L}_{\text{soft}}\supset \frac{g^2_{5}}{2}M\tilde{B}_{1/2}(0)\gl \gl + \text{c.c.}
\ee
$M\tilde{B}_{1/2}(0)$ is the Fourier transform of the Wick contracted two point function fo the gaugino current $j_{\alpha}$, evaluated at zero momenta. For more details of encoding gauge mediated supersymmetry breaking into current correlators we highlight \cite{Meade:2008wd,Komargodski:2008ax,Buican:2008ws,Benakli:2008pg,Intriligator:2008fr,Distler:2008bt,Ooguri:2008ez,Lee:2010kb,Argurio:2010fn,Kang:2010ye}. On the visible fixed point (visible brane of figure \ref{flatfigure}) we locate chiral superfields to represent the supersymmetric standard model matter, whose scalar components we will refer to as an sfermion. Computing the diagrams in figure \ref{flatfigure} as was carried out in \cite{McGarrie:2010kh} for a brane localised chiral superfield, one obtains an sfermion mass given by
\begin{equation}
m_{\tilde{f}}^2=  g_{5}^4  E
\end{equation}
where
\begin{equation}
E= -\! \!\int\! \frac{d^4p}{ (2\pi)^4}\frac {1}{\ell^{2}}\sum_{n, \hat{n}} \!  \frac{(-1)^{n+\hat{n}}}{p^2-(p_{5})^{2}}\frac{p^{2}}{p^2-(\hat{p}_{5})^{2}} [3\tilde{C}_1(p^2/M^2)-4\tilde{C}_{1/2}(p^2/M^2)+\tilde{C}_{0}(p^2/M^2)]. \label{Primeresult}
\end{equation}
For simplicity we have chosen a $U(1)$ model and drop all gauge group labels and Casimirs: these will be reinstated in section  \ref{section:hybrid}  which are the main results of this paper.  Similarly the bulk scalar mass result is 
\begin{equation}
E= -\! \!\int\! \frac{d^4p}{ (2\pi)^4}\sum_{n}(\frac{p}{p^2-p^2_{5}})^2 [3\tilde{C}_1(p^2/M^2)-4\tilde{C}_{1/2}(p^2/M^2)+\tilde{C}_{0}(p^2/M^2)]. 
\end{equation}

\begin{figure}[htcb]
\centering
\includegraphics[scale=0.6]{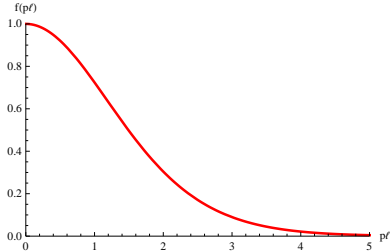}
\caption{A plot of the momentum dependent form factor that suppresses leading order sfermion masses due to bulk mediation of supersymmetry breaking, after a Matsubara summation of all kk modes.  When $p\ell\rightarrow 0 $ one recovers four dimensional general gauge mediation.  When $p\ell \ge 0 $ one obtains screening of the leading order scalar soft mass. }
\label{plot1}
\end{figure}

Using a Matsubara frequency summation of the kk tower, the sfermion mass result may be written
\begin{equation}
E= \! -\!\int\! \frac{d^4p}{ (2\pi)^4} f(p\ell) \frac{1}{p^2} [3\tilde{C}_1(p^2/M^2)-4\tilde{C}_{1/2}(p^2/M^2)+\tilde{C}_{0}(p^2/M^2)] \label{anotherkeyresult}
\end{equation}
This is the four dimensional gauge mediated result \cite{Meade:2008wd} (after rescaling $g^2_{4d}=g^2_{5}/\ell$), multiplied by a momentum dependent form factor 
\be
f(p\ell)=   (\frac{p\ell}{\sinh p\ell})^{2}.
\ee
The form factor is plotted in figure \ref{plot1} and is characteristic of these higher dimensional models.\footnote{Interestingly, form factors of a different kind also arise in Semi-Direct gauge mediation \cite{Argurio:2009ge}.}
\begin{flushleft}
\end{flushleft}
This result has two analytic limits: when the characteristic mass scale of the outer loop $1/\ell$, is larger than the characteristic mass scale $M$ of the current correlators, the form factor will be order 1 and one obtains the four dimensional limit.  This result may be obtained either by setting the form factor to $1$ and evaluating the two loop graphs exactly, as in \cite{Martin:1996zb}, or by starting with the full result, taking an expansion of the current correlators in $M^2/p^2$ and matching the asymptotic behaviour with \cite{Martin:1996zb} accordingly. Conversely, when the characteristic mass scale of the outer loop $1/\ell$, is smaller than the characteristic mass scale $M$ of the hidden sector the form factor suppresses large momentum contributions and one must expand the current correlators in the IR, i.e. in $p^2/M^2$.  This analysis was first carried out in \cite{Mirabelli:1997aj,McGarrie:2010kh}.  To obtain an analytic result in the intermediate regime where $\frac{1}{\ell}\sim M$ requires a different type of expansion than expanding in a ratio of $M$ and $p$. It turns out that for the general case of all kk modes contributing to the mediation of supersymmetry breaking effects this regime cannot be accessed  analytically (it may be obtainable numerically). However for a minimally truncated kk tower (keeping only the massless modes and 1st kk mode of the bulk superfields) analytic results can be obtained \cite{Auzzi:2010xc,Auzzi:2010mb}. 

\section{The minimally truncated model}\label{section:truncated}
\begin{figure}[ht]
\centering
\includegraphics[scale=0.6]{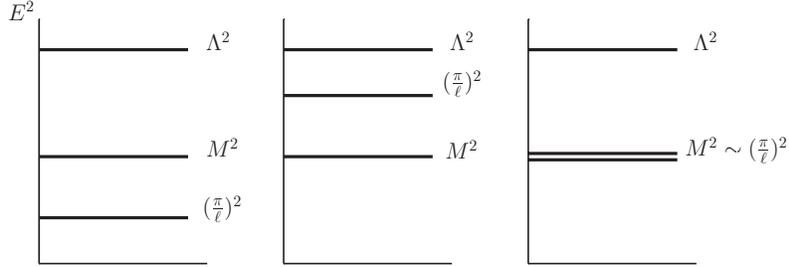}
\caption{Pictorial of the relative mass scales for the two state Kaluza Klein tower for a) gaugino mediation, b) gauge mediation and c) hybrid mediation.  $\Lambda$ is the cutoff of the effective model, $M$ is the characteristic mass scale of the hidden sector or vev of the spurion and the first Kaluza Klein mode is $m_{v}=\frac{\pi}{\ell}$ where $\ell$ is the length of the interval.  We emphasise that when the sfermion masses are screened in the gaugino mediated limit, there are always kk modes below the scale $M$.}
\label{energyscales}
\end{figure}

We have seen in the previous section that all the kk modes of the vector superfield propagate the supersymmetry breaking effects from the hidden to the visible brane.  However, if the mass scales of the model are sufficiently separated, we are at liberty to analyse an effective model in which only the zero mode $m_{0}=0$ and first mode of the kk tower with mass $m_{1}=m_{v}=\frac{\pi}{\ell}$ of vector superfields are part of the spectrum by placing a cutoff $\Lambda$ above these scales, as depicted in figure \ref{energyscales}. As finite truncations of a Kaluza-Klein tower in the vector superfield should be thought of as equivalent in the IR to a finite lattice (de)construction model \cite{McGarrie:2010qr,Csaki:2001em}, the results of this minimal truncation can be related to the minimal (de)construction model explored in \cite{Auzzi:2010xc,Auzzi:2010mb,Sudano:2010vt}. 
\begin{flushleft}
\end{flushleft}
\subsection{Gaugino soft masses}
We start by highlighting the 4 leading soft gaugino mass contributions which are given by
\be 
\mathcal{L}_{\text{soft}}\supset \frac{g^2_{5}}{2}MB_{1/2}(0)(\gl^0 \gl^0 + \gl^1 \gl^0 +\gl^0 \gl^1+ \gl^1 \gl^1) +  \text{c.c.} \label{gauginosoftmass}
\ee
Due to these soft masses, when computing the RG gaugino mediated contributions (as in figure \ref{figure2}) above the mass scale $m_{v}$ this will require evaluating the RG contributions from 6 diagrams.  When running RG equations at energy scales below $m_v$, only the diagram built from $m_{\gl}\gl^0 \gl^0$ contribute and the four dimensional RGE's are sufficient.

\subsection{Sfermion masses}
\begin{figure}[htcb]
\centering
\includegraphics[scale=0.6]{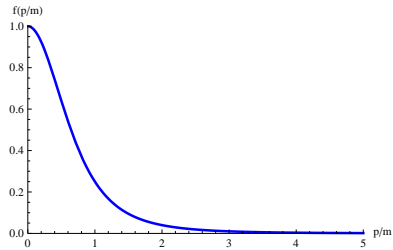}
\caption{A plot of the momentum dependent form factor, for the minimal model.  When $p/m_{v}\rightarrow 0 $ one recovers four dimensional general gauge mediation.  When $p/m_{v} \gg 0 $ one obtains screening of the leading order scalar soft mass.  This model captures the essential features of the all order kk model.}
\label{plot2}
\end{figure}

Supersymmetry breaking effects encoded on the hidden brane are mediated to the visible brane by both modes $\sum_{n=0}^{1}V_{n}(x,y)$. One may write the brane localised sfermion mass summations of \refe{Primeresult} as a product. After a Wick rotation one finds\footnote{We have rescaled the factors of $\ell$ into the coupling}
\begin{equation}
E= -\! \!\int\! \frac{d^4p}{ (2\pi)^4}\frac{1}{p^2}(\frac{m^2_{v}}{p^2+m^2_{v}})^2[3\tilde{C}_1(p^2/M^2)-4\tilde{C}_{1/2}(p^2/M^2)+\tilde{C}_{0}(p^2/M^2)] \label{product}
\end{equation}
with 
\be 
f(p/m_{v})=\left(1/(\frac{p^2}{m^2_{v}}+1)\right)^2.
\ee
The form factor is plotted in figure \ref{plot2} and captures the essential screening behaviour of the all order model plotted in figure \ref{plot1}.  This result has also been obtained for the (de)constructed version of this model \cite{McGarrie:2010qr,Auzzi:2010xc,Sudano:2010vt}. 
\begin{flushleft}
\end{flushleft}
\subsection{Bulk scalar masses}
In \cite{McGarrie:2010kh} the positive parity bulk scalar zero mode masses $m^2_{H_{0}}$ were computed.  For the two state model it is given by  
\begin{equation}
m^2_{H_{0}}=  g_{4}^4   D
\end{equation}
where
\begin{equation}
D= -\! \!\int\! \frac{d^4p}{ (2\pi)^4}\sum_{n=0}^{1}(\frac{p}{p^2-m^2_{n}})^2 [3\tilde{C}_1(p^2/M^2)-4\tilde{C}_{1/2}(p^2/M^2)+\tilde{C}_{0}(p^2/M^2)]. \label{hyperscalar}
\end{equation}
This can be rewritten as (we Wick rotated and then manipulated)
\begin{equation}
D= -\! \!\int\! \frac{d^4p}{ (2\pi)^4}(\frac{1}{p^2} +\frac{1}{p^2+m^2_{v}}- \frac{m^2_{v}}{[p^2+m^2_{v}]^2}) [3\tilde{C}_1(p^2/M^2)-4\tilde{C}_{1/2}(p^2/M^2)+\tilde{C}_{0}(p^2/M^2)]. \label{Thekeyequation}
\end{equation}
We will show in section  \ref{section:hybrid} how this soft mass may be analytically determined for a specific hidden sector and specified currents.  In the next section we will show that this mass formula will determine the soft masses of linking fields in the minimal gaugino mediation model.
\subsection{Minimal gaugino mediation}
\begin{figure}[ht]
\centering
\includegraphics[scale=0.6]{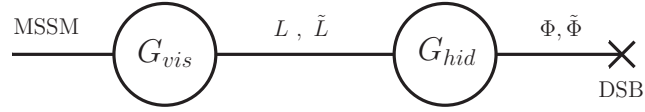}
\caption{Minimal gaugino mediation. MSSM matter located at the lattice site $G_{vis}$ corresponds to matter on the visible brane in figure 1. Similarly the fields $\Phi,\tilde{\Phi}$ located at $G_{hid}$ represents messengers located on the hidden brane.  The lattice linking fields $L$ and $\tilde{L}$ are bifundamental and antibifundamental respectively and corresponds to bulk hypermultiplets of figure 1. }
\label{model}
\end{figure}

The minimally truncated five dimensional model we have been describing so far can be related, in the IR, to the  minimal gaugino mediation model pictured in figure \ref{model} \cite{Auzzi:2010xc,Sudano:2010vt,Auzzi:2010mb}. In this section we would like to further clarify this connection.  We will give leading order soft mass formulas for the linking fields and show that this soft mass is the same soft mass contribution as for a bulk scalar as in the five dimensional model \refe{hyperscalar}.
\begin{flushleft}
\end{flushleft}
\subsection{Linking scalar soft masses}
For a general deconstructed model of $N$ lattice sites labelled from $i=0,...,N-1$ (for more details see \cite{McGarrie:2010qr,Csaki:2001em}), the leading order sfermion mass contribution from a supersymmetry breaking lattice site at $N-1$ to any scalar located at another lattice site $k$ is given by 
\begin{equation}
E\!\!=\!\! -\! \!\int\! \frac{d^4p}{ (2\pi)^4}p^{2}(\braket{p^2;k,N-1})^2
[3\tilde{C}_1(p^2/M^2)-4\tilde{C}_{1/2}(p^2/M^2)+\tilde{C}_{0}(p^2/M^2)]
\end{equation}
where the bulk propagator is given by 
\be
\braket{p^2;k,l} =\frac{1}{N}\sum_{j=0}^{N-1}e^{i(2\pi k j)/N}e^{i(2\pi \ell j)/N}\frac{1}{p^2+m^2_{j}} \label{propagator2}
\ee
with mass eigenstates given by 
\be
m^2_{k}=8g^2 v^2 \sin^2 (\frac{k\pi}{N}) \phantom{AAAAA} k=0, ..., N-1.
\ee
It is important to stress that the lattice eigenstates are not mass eigenstates of the system and it is in this change of basis that the propagator is derived.  The kinetic terms for the bifundamental chiral superfields of the minimal model are given by
\be \delta {\cal L} = \int d^2\theta d^2\bar\theta \left(L^\dag
e^{2 g V_{0} -2 g V_{1} } L + \ti L^\dag e^{-2 g V_0+2 g V_{1} }
\ti L \right)  \label{linkingfields} \ee
where for simplicity we have taken $g_0=g_1=g$.  We see that for the minimal model of figure \ref{model} the linking fields will each pick up two leading order soft masses, being bifundamental.  The resulting soft mass for $m^2_{L} L^\dagger L$  is given by $m^2_{L}=\sum_{k=0}^{1}m^2_{k,L}$ where
\begin{equation}
m^2_{0,L}\!\!=\!\! -g^4\! \!\int\! \frac{d^4p}{ (2\pi)^4}p^{2}(\braket{p^2;0,1})^2
[3\tilde{C}_1(p^2/M^2)-4\tilde{C}_{1/2}(p^2/M^2)+\tilde{C}_{0}(p^2/M^2)],
\end{equation}
\begin{equation}
m^2_{1,L}\!\!=\!\! -g^4\! \!\int\! \frac{d^4p}{ (2\pi)^4}p^{2}(\braket{p^2;1,1})^2
[3\tilde{C}_1(p^2/M^2)-4\tilde{C}_{1/2}(p^2/M^2)+\tilde{C}_{0}(p^2/M^2)].
\end{equation}
Adding these contributions, one obtains the bulk scalar mass result \refe{hyperscalar}.  The mass eigenstates are then given by the full mass matrix including the kk masses and all linking fields of the same representation.

\section{Hybrid gauge mediation}\label{section:hybrid}
\begin{figure}[ht]
\centering
\includegraphics[scale=0.6]{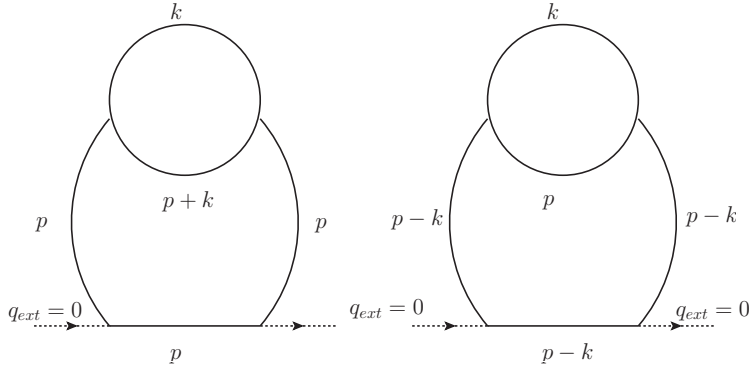}
\caption{A figure representing the labelling of momenta in the typical two-loop diagrams contributing to sfermion masses at leading order.  The inner loop has a characteristic mass scale $M$ and is typically a loop inside the current correlator, when the hidden sector has a perturbative description. The outer loop has a characteristic mass scale $m_{v}$.  The first case of labelling momenta is typical for GGM, however the mass scales $M$ and $m_{v}$ do not mix within each integral on $k$ or $p$ momenta and one must expand current correlators in a ratio of $M^2$ and $p^2$ to obtain an analytic limit. In the second case, the mass scales mix in either integral on momenta $k$ and $p$ and one may expand in a ratio of $m_{v}/M$.}
\label{labelmomentum}
\end{figure}

So far, we have given quite general soft mass expressions for gauginos, visible brane localised scalars and bulk positive parity scalars for the minimal model.  In this section we will specify the supersymmetry breaking hidden sector to be a generalised messenger sector coupled to a supersymmetry breaking spurion.  Specifying the hidden sector specifies the currents of the hidden sector and we may then use known expressions for these currents.  The relevant diagrams are typically two loop diagrams whose momentum integrals can be found in  \cite{Meade:2008wd,McGarrie:2010kh,Martin:1996zb,Marques:2009yu} and whose momenta is typically labelled as in figure \ref{labelmomentum}.  It is quite straightforward to shift momenta of these two loop diagrams and then apply the general expressions for massive two-loop Feynman diagrams, which are analytically solvable when the Mandelstam variables vanish \cite{Ghinculov:1994sd,vanderBij:1983bw,Martin:1996zb,Marques:2009yu}. 
\begin{flushleft}
\end{flushleft}
\subsection{Generalised messenger sector}\label{section:general}
In this section we give a concrete description of matter content of the SUSY breaking sector located on the hidden brane (hidden lattice site), following the construction of \cite{Martin:1996zb}. We consider sets of $N$ chiral superfield messengers  $\Phi_{i},\tilde{\Phi}_{i}$ in the vector like representation of the lattice gauge group, coupled to a SUSY breaking spurion $X= M + \theta^2 F $. Generalisations to arbitrary hidden sectors are a straightforward application of the results of \cite{Marques:2009yu,McGarrie:2010kh}.  The superpotential is 
\be W  =X \eta_{i}\
\Phi_i \ti \Phi_i \label{superpotential2}\ee
In principle $\eta_{ij}$ is a generic matrix which may be diagonalised to its eigenvalues $\eta_{i}$ \cite{Martin:1996zb}.  The messengers will couple to the bulk vector superfield as
\be \delta {\cal L} = \int d^2\theta d^2\bar\theta \left(\Phi^\dag_i
e^{2 g V^a T^a} \Phi_i + \ti\Phi^\dag_i e^{-2 g V^a T^a}
\ti\Phi_i\right) + \left(\int d^2\theta\  W +
c.c.\right) \label{hiddensector} \ee
We can extract the multiplet of currents from the kinetic terms in the above Lagrangian. The current correlators can then be computed and their results can be found in \cite{Meade:2008wd,McGarrie:2010kh,Marques:2009yu}.  We will use the result of these current correlators to determine the gaugino massses, sfermion masses on the visible brane and the (positive parity) bulk scalar soft mass.
\subsection{Gaugino masses}
The current correlator in \refe{gauginosoftmass} may be evaluated using the currents found in \cite{Meade:2008wd,McGarrie:2010kh,Martin:1996zb,Marques:2009yu} for the general messenger sector described above.   The zero mode gaugino mass is found to be 
\be 
m^r_{\gl_{0} }= \frac{\alpha_{r}}{4\pi} \Lambda_{G}\ , \ \ \ \  \Lambda_{G}=\sum_{i=1}^{N}[\frac{d_{r}(i) F}{M}g(x_{i})]
\label{gauginomass}
\ee
The label $r=1,2,3$ refers to the gauge groups $U(1),SU(2),SU(3)$, $d_{r}(i)$ is the Dynkin index of the representation of $\Phi_{i},\tilde{\Phi}_{i}$ and
\be 
g(x)= \frac{(1-x)\log(1-x)+(1+x)\log(1+x)}{ x^2}
 \ee
where  $x_{i}=\frac{F}{\eta_{i}M^2}$.  $g(x)\sim 1$ for small $x$ \cite{Martin:1996zb}.

\subsection{sfermion masses}
Using the result of \cite{Auzzi:2010mb} one finds the sfermion masses on the visible brane is given by 
\be
 m^2_{\tilde{f}}=2\left(\frac{F}{M}\right)^2 \sum_{r}(\frac{\alpha_{r}}{4\pi})^2 c(\tilde{f},r)\sum_{i} d_{r}(i)S(x_{i},y_{i})
\ee
with $y_{i}=m_{v}/\eta_{i}M$, where $c(\tilde{f},r)$ is the quadratic Casimir of the gauge group $r$ for the MSSM scalar of representation $\tilde{f}$.  $S(x,y)$ is given in the appendix.

\subsection{Bulk scalar masses}
The positive parity bulk scalar (linking scalar) mass, is given by 
\be
 m^2_{\tilde{h}}=2\left(\frac{F}{M}\right)^2 \sum_{r}(\frac{\alpha_{r}}{4\pi})^2 c(\tilde{h},r)\sum_{i} d_{r}(i)G(x_{i},y_{i})
\ee
where $G(x,y)$ is given in the appendix.

\begin{flushleft}
\end{flushleft}
\begin{figure}[ht]
\centering
\includegraphics[scale=0.6]{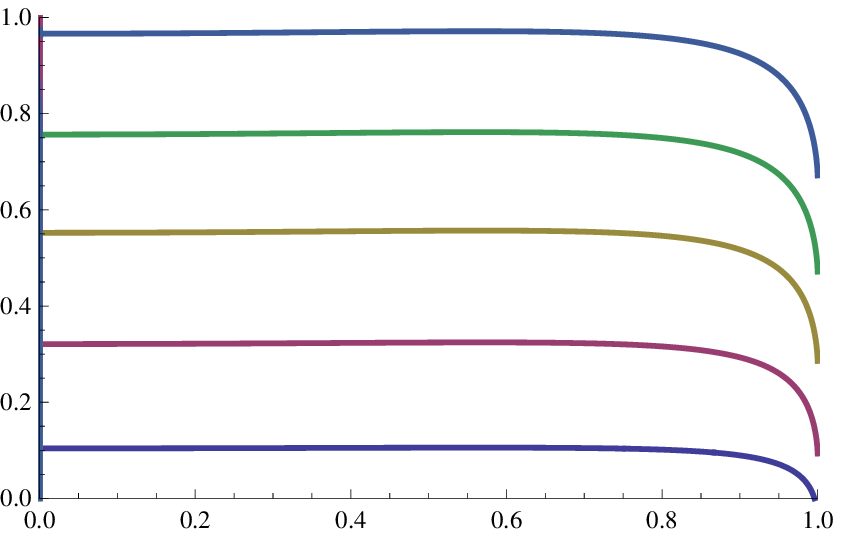}
\includegraphics[scale=0.6]{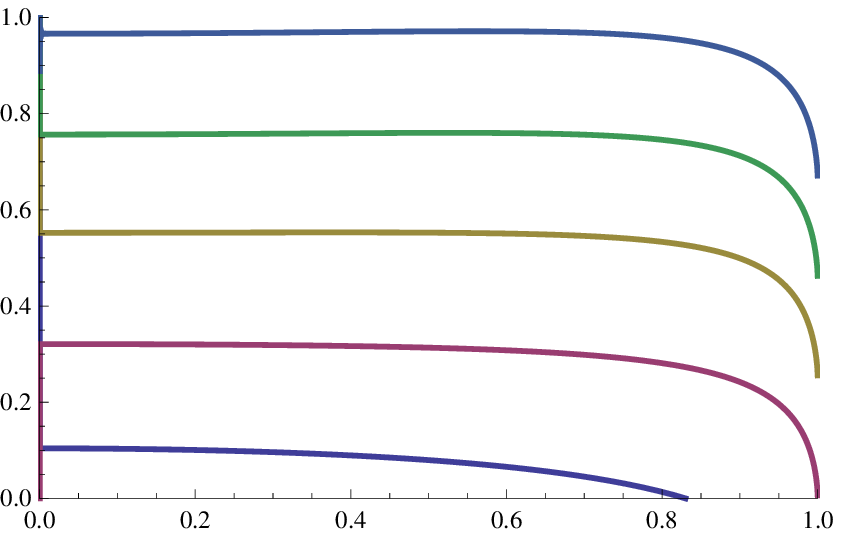}
\caption{On the left is a figure of $S(x,y)$ plotted along the $x$ axis going from bottom to top are $y=1,2.5,5,10,50.$  On the right is a plot of $G(x,y)$ for the same values of $y$. The right plot shows that for small values of $y$, $G(x,y)$ becomes tachyonic as $x$ approaches $1$, far more quickly than $S(x,y)$.}
\label{suppressionplots}
\end{figure}

\begin{flushleft}
\end{flushleft}
\begin{figure}[ht]
\centering
\includegraphics[scale=0.6]{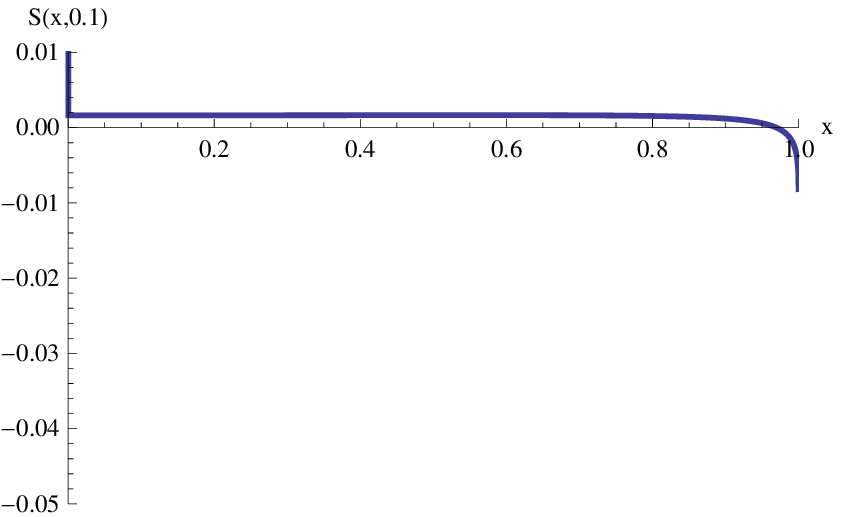}
\includegraphics[scale=0.6]{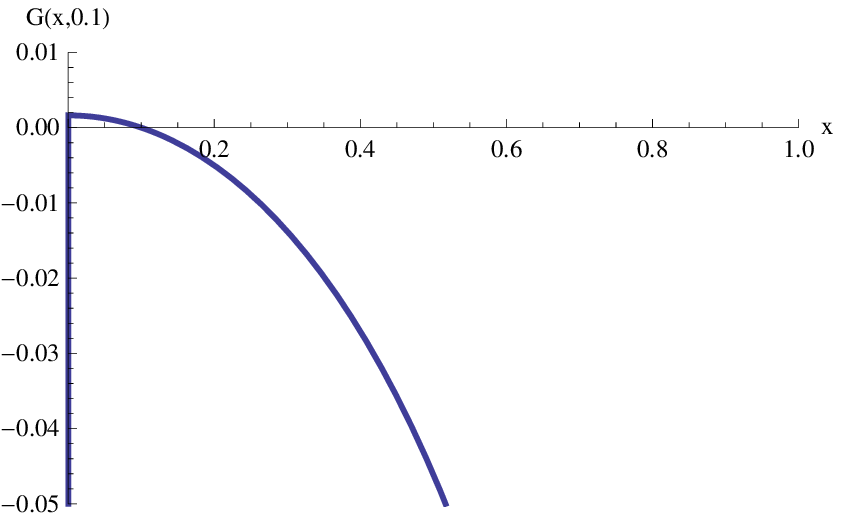}
\caption{On the left is a plot of $S(x,0.1)$ and on the right $G(x,0.1)$ along the x axis.  On the right plot we see the function becoming tachyonic far more quickly than on the left.}
\label{tachy}
\end{figure}

\begin{flushleft}
\end{flushleft}
\begin{figure}[ht]
\centering
\includegraphics[scale=0.6]{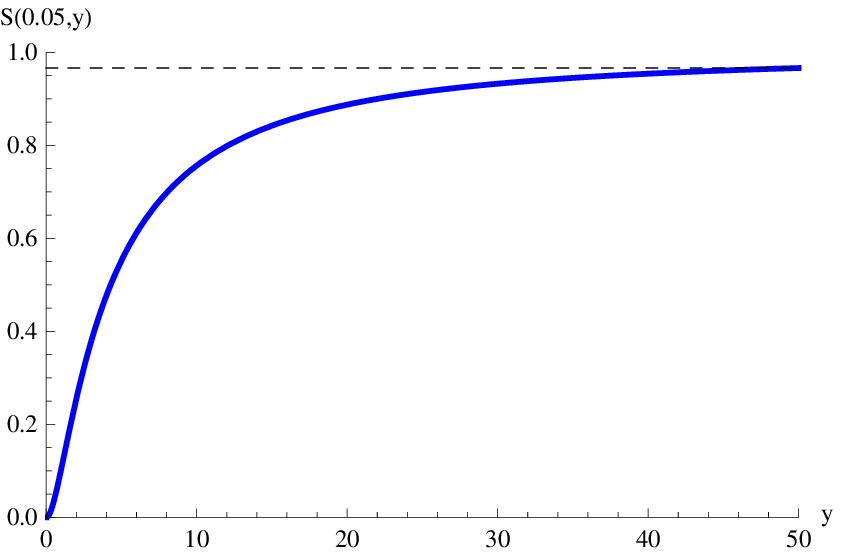}
\includegraphics[scale=0.6]{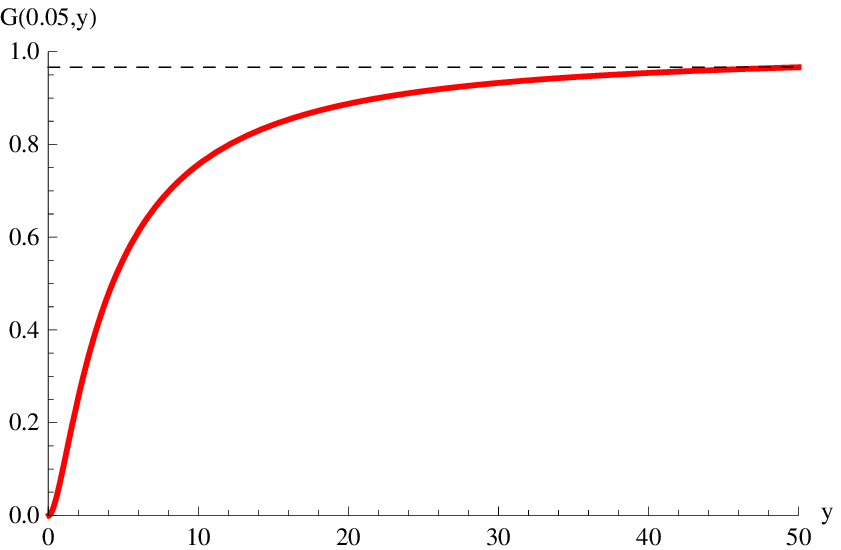}
\caption{On the left is a figure of $S(0.05,y)$ and on the right $G(0.05,y)$.  For small $y$ the functions are screened and scale as $\sim y^2$ and interpolates through the hybrid regime to large $\sim y^0$: the 4d limit.  }
\label{contour}
\end{figure}

In figure \ref{suppressionplots}, \ref{tachy} and \ref{contour}  we compare plots of the functions $S(x,y)$ for sfermion masses and $G(x,y)$ for scalar masses. The plots may become tachyonic for some regimes of the parameter space and both plots have similar behaviour with regimes of strong screening of the masses.  Interestingly, these techniques to obtain analytic results may also be applicable for models with gauge messengers \cite{Intriligator:2008fr,Intriligator:2010be,Matos:2010ie}.

\section{Subleading contributions to general gauge mediation}\label{section:subleading}
In this section we would like to discuss the computation of some subleading diagrams \cite{Chacko:1999mi,Kaplan:1999ac,Csaki:2001em}.  We focus on the subleading soft mass ($p_{ext}=0$) correction in figure \ref{figure2}.  In the all order kk model \cite{McGarrie:2010kh}  its contribution is given by
\be
\delta m_{\tilde{f}}^2= \sum_r c_2(\tilde{f};r) \frac{g^6_{5}}{2\ell^3}\int \frac{d^4p }{(2\pi)^4}p^2\sum_{n,\hat{n},\hat{\hat{n}}}\frac{(-1)^{n+\hat{n}}}{p^2+p^2_{5}}\frac{M\tilde{B}_{1/2}(p^2/M^2)}{p^2+\hat{p}^2_{5}}\frac{M\tilde{B}_{1/2}(p^2/M^2)}{p^2+\hat{\hat{p}}^2_{5}}.
\ee
In general, this integral is divergent due to the brane to same brane propagator on the hidden brane that connects the double mass insertions.   The four dimensional limit, when $M\ll \frac{1}{\ell}$ i.e. $\ell$ is small one finds 
\be
 \delta m_{\tilde{f}}^2= \sum_r c_2(\tilde{f};r)\frac{g^6_{4d}}{2} \int \frac{d^4p }{(2\pi)^4}\frac{1}{p^4}(M\tilde{B}_{1/2}(p^2/M^2))^2.
\ee
In the limit $\frac{1}{\ell}\ll M $ one may carry out a Matsubara summation and finds 
\be
\delta m_{\tilde{f}}^2= \sum_r c_2(\tilde{f};r)\frac{g^6_{5}}{2}\int \frac{d^4p }{(2\pi)^4}\frac{\left(M\tilde{B}_{1/2}(0)\right)^2}{p \sinh^2 (p\ell)\tanh (p\ell) }
\ee
where in this limit 
\be 
\lim_{\frac{p^2}{M^2}\rightarrow 0}M\tilde{B}_{1/2}(p^2/M^2)=M\tilde{B}_{1/2}(0).
\ee
The integral is IR divergent and must be regulated by use of the effective cutoff $\Lambda$. We choose  
\be
\delta m_{\tilde{f}}^2= \sum_r c_2(\tilde{f};r)\frac{g^6_{4}}{2}\left(M\tilde{B}_{1/2}(0)\right)^2 \int_{0}^{\infty} \frac{2 \pi^2 dy }{(2\pi)^4}[\frac{y^2}{ \sinh^2 (y)\tanh (y) }-\frac{e^{-\Lambda y}}{y}]
\ee
\be
\delta m_{\tilde{f}}^2= \sum_r c_2(\tilde{f};r)\frac{g^6_{4}}{16\pi^2}\left(M\tilde{B}_{1/2}(0)\right)^2[3/2 + \gamma +\log \Lambda/2 ],
\ee
which for a generalised messenger sector, one may use \refe{gauginomass}. Schematically, for a messenger sector, one can compare the leading order and subleading contribution
\be
 m^2_{\tilde{f}}\sim (\frac{\alpha}{4\pi})^2 \Lambda^2_{S}\frac{1}{(M \ell)^2}+ (\frac{\alpha}{4\pi})^3 \Lambda^{2}_{G}
\ee
such that it is not always clear which term is truly leading order. $\Lambda_S$ is the four dimensional scalar mass scale and $\Lambda_G$ is the gaugino mass scale.  When $\Lambda_{S}\sim \Lambda_{G}$ comparing $\alpha/4\pi$ versus $1/(M\ell)^2$ may be sufficient, however models with an approximate R-symmetry may also suppress $\Lambda_{G}$.  It is worth emphasising that for the generalised messenger sector discussed in this paper and for typical values of $M$, $\ell$ and $\alpha$, the double mass insertion in figure \ref{figure2} is most likely to be the largest contribution to sfermion masses, however for ISS-like models \cite{McGarrie:2010kh,Green:2010ww} where the gaugino mass is suppressed due to an R-symmetry, figure \ref{flatfigure} is the leading contribution.   It is also useful to note that the subleading diagram may act as a useful bound, at the high scale $M$, on the ratio of masses:
\be
\frac{m^2_{\lambda} }{m^2_{\tilde{f}}}\lesssim (\frac{4\pi}{\alpha})\sim 300 
\ee
where in the last line we took $\alpha(M_{GUT})$=0.04.

We briefly comment on warped general gauge mediation \cite{McGarrie:2010yk,Abel:2010vb,Bouchart:2011va} in which the AdS warp factor controls the mass of the kk modes $m_{n}\sim n\pi k e^{-k\ell}$ rather than $\ell$, and in the regime $k\ll M$ one finds 
\be
\delta m_{\tilde{f}}^2= \sum_r c_2(\tilde{f};r)\frac{g^6_{5d}}{16\pi^2 \ell^3}(k\ell)^3 e^{-k\ell} \left(\hat{M}\tilde{B}_{1/2}(0)\right)^2[3/2 + \gamma +\log \Lambda/2 ].
\ee
which is hierarchically suppressed by $e^{-k\ell}$ as expected. For this result one may use \refe{gauginomass} with $F\rightarrow \hat{F}=e^{-2k\ell}F$ and $M\rightarrow \hat{M}=e^{-k\ell}M$. As a result,  when $k\ll M$, these models have a rather different bound
\be
\frac{m^2_{\lambda} }{m^2_{\tilde{f}}}\lesssim (\frac{4\pi}{\alpha})(k\ell)^2 e^{k\ell}
\ee
and it seems we can have an exponential hierarchy between the two soft mass scales.    

 For the minimal (flat) two state model one obtains (after Wick rotation)
\be
 \delta m_{\tilde{f}}^2= \sum_r c_2(\tilde{f};r)g^6_{4d} \int \frac{d^4p }{(2\pi)^4}\frac{m^4_{v}}{p^4 (p^2+m^2_{v})^2}(M\tilde{B}_{1/2}(p^2/M^2))^2
\ee
which carries the same momentum dependent form factor as figure \ref{plot2}.  Keeping the full momentum dependence of the mass insertion, this is a three loop contribution and unfortunately cannot be calculated analytically by the same techniques that applied to the two loop leading order contributions of the previous section.  Additionally, there are three loop contributions to sfermion masses from bulk scalar masses, as has been commented before \cite{Green:2010ww}.  It would certainly be interesting to apply numerical techniques to these superficially subleading diagrams (order $g^6$), keeping their full momentum dependence, to better understand their role in this model.

\section{Summary and conclusion}\label{conclusion}
In this paper we have outlined a concrete framework in which higher dimensional models of gauge mediated supersymmetry breaking may obtain a hybrid between gauge and gaugino mediation.  In particular we have shown that the leading order sfermion and bulk scalar masses, being massive two-loop diagrams with zero external momenta, are analytically solvable for this model which allows for a determination of this mass contribution in the hybrid regime that $M\sim m_{v}$ as was first pointed out in \cite{Auzzi:2010mb}.

Three loop contributions, particularly those mention in this paper should be explored numerically to determine their effect in the different regimes.  One might be interested to explore this minimal model for the warped case \cite{McGarrie:2010yk}, although as the first kk modes are often localised towards the IR brane, this mode may have little affect on the mediation of supersymmetry breaking.  We think it worthwhile to apply this minimal two state model to brane to brane supergravity mediation, in particular because that model also solves problems with flavour changing neutral currents and may generate less suppressed soft masses as usually encountered in the full all order kk mode calculations.
\paragraph{Acknowledgments} 
I would like to thank Steven Thomas, Rodolfo Russo, Daniel C. Thompson, Zohar Komargodski, Roberto Auzzi and the organisers and attendees of the Gauge mediated supersymmetry breaking workshop at the Vrije Universiteit Brussel.  I am funded by STFC.
\appendix

\section{Evaluation of the bulk scalar soft mass integrals}\label{appendixA}
In this appendix we will evaluate some integrals relating to the leading order bulk scalar soft mass with a generalised messenger sector of section  \ref{section:general}.  This is a different calculation to the computation of the brane localised scalar soft mass found in \cite{Auzzi:2010mb}, however the techniques are the same and we will use and review those techniques here. The original and more complete references are \cite{Martin:1996zb,Ghinculov:1994sd,vanderBij:1983bw,Marques:2009yu}.

First we define the notation
\be
\langle m_{11}, \ldots, m_{1 n_1} | m_{21} , \ldots , m_{2 n_2} | m_{31},
 \ldots, m_{3 n_3} \rangle
\ee
\be
= \int \frac{d^d k}{\pi^{d/2}}  \frac{d^d q}{\pi^{d/2}} \prod_{i=1}^{n_1}
\prod_{j=1}^{n_2} \prod_{l=1}^{n_3} \frac{1}{k^2+m_{1i}^2}
 \frac{1}{q^2+m_{2j}^2}  \frac{1}{(k-q)^2+m_{3l}^2} \, . \nonumber
\ee
The bulk scalar soft mass for the minimal model is given by \refe{Thekeyequation} and is a sum of three terms.  The first term is the four dimensional soft mass result given by taking $y\rightarrow \infty$ in $S(x,y)$, which gives
\be 
S(x,\infty)=\frac{s_{0}}{2x^2}
\ee
with $s_{0}$ defined below. The second term is

\be
(+ g^4 /(4 \pi)^d) \left( - \langle m_+|m_+|m_v \rangle  -  \langle m_-|m_-|m_v  \rangle \right.
\ee
\be \left.
-4 \langle m_f|m_f|m_v  \rangle  -2  \langle m_+|m_-|m_v  \rangle + 4  \langle m_+|m_f|m_v  \rangle  \right.
\nonumber\ee
\be
\left.  +4 \langle m_-|m_f|m_v  \rangle -4 m_+^2 \langle m_+|m_+|0,m_v  \rangle \right.
\nonumber\ee
\be \left. - 4 m_-^2  \langle m_-|m_-|0,m_v \rangle +  8 m_f^2 \langle m_f|m_f|0,m_v  \rangle \right.
\nonumber\ee
\be \left. +4(m_+^2 -m_f^2)   \langle m_+|m_f|0,m_v \rangle  +
 4(m_-^2 -m_f^2)   \langle m_-|m_f|0,m_v  \rangle
\right) \, .
\nonumber\ee
This result is obtained by removing one massless and one $m_{v}$ entry in each term in \cite{Auzzi:2010mb}. Using the regulator $m_{\epsilon}$ for the massless propagator  and starting from this result, one applies partial fractions 
\be 
\frac{1}{[(p+k)^2-m^2_{1}][(p+k)^2-m^2_{2}]}=\frac{1}{m^2_{1}-m^2_{2}}\left[\frac{1}{(p+k)^2-m^2_{1}} -\frac{1}{(p+k)^2-m^2_{2}}\right]
\ee
to the last 5 terms such that all integrals are of the same form. 
The third term is given by evaluating 
\be
(-m^2_{v} g^4 /(4 \pi)^d) \left( - \langle m_+|m_+|m_v,m_v \rangle  -  \langle m_-|m_-|m_v,m_v  \rangle \right.
\ee
\be \left.
-4 \langle m_f|m_f|m_v,m_v  \rangle  -2  \langle m_+|m_-|m_v,m_v  \rangle + 4  \langle m_+|m_f|m_v,m_v  \rangle  \right.
\nonumber\ee
\be
\left.  +4 \langle m_-|m_f|m_v,m_v  \rangle -4 m_+^2 \langle m_+|m_+|0,m_v,m_v  \rangle \right.
\nonumber\ee
\be \left. - 4 m_-^2  \langle m_-|m_-|0,m_v,m_v  \rangle +  8 m_f^2 \langle m_f|m_f|0,m_v,m_v  \rangle \right.
\nonumber\ee
\be \left. +4(m_+^2 -m_f^2)   \langle m_+|m_f|0,m_v,m_v  \rangle  +
 4(m_-^2 -m_f^2)   \langle m_-|m_f|0,m_v,m_v  \rangle
\right) \, .
\nonumber\ee
This result is obtained by removing a massless entry in each term in \cite{Auzzi:2010mb}. The symbolic manipulations are straightforward applications Mathematica by repeated use of ``Rules''.  We apply 
\be
 \langle m_a | m_b | 0, m_v, m_v  \rangle = \frac{  \langle m_a | m_b | 0 \rangle -   \langle m_a | m_b | m_v \rangle }{m_v^4}
-\frac{ \langle m_a | m_b | m_v , m_v  \rangle}{m_v^2}  \, ,
\ee
and then apply  ($d=4-2 \epsilon$),
\be
 \langle  m_0|m_1|m_2  \rangle=
\frac{1}{-1+2 \epsilon} \left(
m_0^2    \langle  m_0,m_0|m_1|m_2  \rangle  \right.
\ee
\be \left. + m_1^2   \langle m_1,m_1|m_0|m_2  \rangle+
m_2^2  \langle  m_2,m_2|m_0|m_1  \rangle
\right) \, .\nonumber
\ee
This reduces to just the first two terms on the right hand side when $m_2=m_{\epsilon}=0$.  One may also use 
\be 
\langle m_0 | m_1 | m_2, m_2  \rangle= \langle m_2 , m_2 | m_0| m_1 \rangle.
\ee
All the terms are now expressed in terms of the basic object
\be
\langle m_0,m_0|m_1|m_2 \rangle=
\frac{1}{2 \epsilon^2}+\frac{1/2-\gamma - \log m_0^2}{\epsilon}
\ee
\be 
+\gamma^2-\gamma +\frac{\pi^2}{12}
+(2 \gamma -1) \log m_0^2 + \log^2 m_0^2 -\frac{1}{2}+ h(a,b) \, .
\ee
The function $h$ is given by the integral  \cite{vanderBij:1983bw}:
\be
h(a,b)= \int_0^1 dx  \left( 1+ {\rm Li}_2 (1-\mu^2) -\frac{\mu^2}{1-\mu^2}  \log \mu^2 \right) \, 
\ee
The dilogarithm is defined as ${\rm Li}_2(x)=-\int_0^1 \frac{dt}{t}\log(1-xt)$  with
\be
\mu^2=\frac{a x + b(1-x)}{x(1-x)} \, \ \ , \ \   a=m_1^2/m_0^2   \ \ , \ \ b=m_2^2/m_0^2.
\ee
It is useful to first evaluate the terms with massless propagators, whereby the function $h$  simplifies to  $h(0,b)=1+\rm{Li}_2 (1-b)$ and has a symmetry $h(b,0)$=$h(0,b)$.   Then for the terms with entirely massive propagators, the analytic expression for $h$ is used to obtain the plots:
\be
h(a,b)=1-\frac{\log a \log b}{2} -\frac{a+b-1}{\sqrt{\Delta}} \left( {\rm Li}_2 \left (-\frac{u_2}{v_1} \right) + {\rm Li}_2 \left (-\frac{v_2}{u_1} \right)
\right.
\ee
\be
\left. + \frac{1}{4} \log^2 \frac{u_2}{v_1}  +  \frac{1}{4} \log^2 \frac{v_2}{u_1} +    \frac{1}{4} \log^2 \frac{u_1}{v_1} -    \frac{1}{4} \log^2 \frac{u_2}{v_2} + \frac{\pi^2}{6} \right) \, ,
\ee
where
\be
\Delta= 1-2 (a+b) +(a-b)^2 \, ,  \qquad u_{1,2}= \frac{1+b-a \pm \sqrt{\Delta}}{2} \, ,
\ee
\be v_{1,2}=\frac{1-b+ a \pm \sqrt{\Delta} }{2} \, .
\ee
The final result is that 
\be
m^2_{\tilde{h}}= 4(\frac{\alpha}{4\pi})^2(\frac{F}{M})^2 G(x,y)
\ee
where 
\be
G(x,y)= \frac{1}{2x^2} (s_{0}+ \frac{t_{1}+t_{2}}{y^2}+t_{3}+y^2 t_{4} )+ (x \rightarrow -x)
\ee
\be
s_0=2(1+x) \left( \log (1+x) -2 {\rm Li}_2  \left( \frac{ x}{1+x}\right)
+\frac{1}{2} {\rm Li}_2 \left( \frac{2 x}{1+x}\right) \right)  ,
\ee
\be
t_{1}=-4x^2-2x(1+x) \log (1+x)-x^2 {\rm Li}_2 \left(x^2\right) \nonumber
\ee
\bea
t_{2}&=-8h\left(1,y^2\right) +8(1+x)^2 h\left(1,\frac{y^2}{1+x}\right)
\nonumber
\\ &-4x h\left(1+x,y^2\right) -4x(1+x)h\left(\frac{1}{1+x},\frac{y^2}{1+x}\right) \nonumber
\eea
\bea
t_{3}&=2h\left(1,y^2 \right)+(1+x)h\left(1,\frac{y^2}{1+x}\right)-2h\left(1+x,y^2 \right)+(1+x)h\left(\frac{1-x}{1+x},\frac{y^2}{1+x}\right) \nonumber \\&  -2 h\left(\frac{1}{y^2},\frac{1}{y^2}\right) -2(1+x) h\left(\frac{1}{1+x},\frac{y^2}{1+x}\right)+ 2(1+x) h\left(\frac{1+x}{y^2},\frac{1+x}{y^2} \right)-2x h\left(\frac{1+x}{y^2},\frac{1}{y^2}\right)\nonumber
\eea
\be
t_{4}= 2h\left(\frac{1}{y^2},\frac{1}{y^2}\right)-4h\left(\frac{1+x}{y^2},\frac{1}{y^2}\right)+h\left(\frac{1+x}{y^2},\frac{1+x}{y^2}\right)+h\left(\frac{1+x}{y^2},\frac{1-x}{y^2} \right) .\nonumber
\ee
As a consistency check we also computed the result of  \cite{Auzzi:2010mb} after applying further rules presented in that paper, which gives an analytic expression for $s(x,y)$:
\be
s(x,y)=\frac{1}{2x^2}\left(s_0 +\frac{s_1+ s_2}{y^2} + s_3 +s_4 +s_5 \right)
 + \, (x \rightarrow -x)\, , 
\ee
where
\be
s_0=2(1+x) \left( \log (1+x) -2 {\rm Li}_2  \left( \frac{ x}{1+x}\right)
+\frac{1}{2} {\rm Li}_2 \left( \frac{2 x}{1+x}\right) \right)   \, , 
\ee

\be
s_1=- 4 x^2   - 2 x(1+x) \log^2(1+x) - x^2 \, {\rm Li}_2(x^2) \, ,
\nonumber\ee
\be
s_2=8 \left(1+x\right)^2 h\left(\frac{y^2}{1+x},1\right)-4 x \left(1+x\right) h\left(\frac{y^2}{1+x},\frac{1}{1+x}\right)
\nonumber\ee
  \be
   -4 x    h\left(y^2,1+x \right)-8 h\left(y^2,1\right) \, ,
\nonumber\ee
\be
s_3= -2 h\left(\frac{1}{y^2},\frac{1}{y^2}\right)
-2 x \,   h\left(\frac{1+x}{y^2},\frac{1}{y^2}\right) +
2(1+ x) h\left(\frac{1+x}{y^2},\frac{1+x}{y^2}\right) \, ,
\nonumber \ee
\be
s_4=(1+x) \left(  2  h\left(\frac{y^2}{1+x},\frac{1}{1+x}\right)
 - h\left(\frac{y^2}{1+x},1\right)- h\left(\frac{y^2}{1+x},\frac{1-x}{1+x}\right)  \right) \, ,
\nonumber \ee
\be
s_5= 2 h\left(y^2,1+x\right)-2 h\left(y^2,1\right) \, . \nonumber
\ee

\begin{thebibliography}{10}

\bibitem{Meade:2008wd}
P.~Meade, N.~Seiberg, and D.~Shih, {\it {General Gauge Mediation}},  {\em Prog.
  Theor. Phys. Suppl.} {\bf 177} (2009) 143--158,
  [\href{http://xxx.lanl.gov/abs/0801.3278}{{\tt arXiv:0801.3278}}].

\bibitem{Mirabelli:1997aj}
E.~A. Mirabelli and M.~E. Peskin, {\it {Transmission of supersymmetry breaking
  from a 4- dimensional boundary}},  {\em Phys. Rev.} {\bf D58} (1998) 065002,
  [\href{http://xxx.lanl.gov/abs/hep-th/9712214}{{\tt hep-th/9712214}}].

\bibitem{McGarrie:2010kh}
M.~McGarrie and R.~Russo, {\it {General Gauge Mediation in 5D}},  {\em Phys.
  Rev.} {\bf D82} (2010) 035001, [\href{http://xxx.lanl.gov/abs/1004.3305}{{\tt
  arXiv:1004.3305}}].

\bibitem{McGarrie:2010qr}
M.~McGarrie, {\it {General Gauge Mediation and Deconstruction}},  {\em JHEP}
  {\bf 11} (2010) 152, [\href{http://xxx.lanl.gov/abs/1009.0012}{{\tt
  arXiv:1009.0012}}].

\bibitem{McGarrie:2010yk}
M.~McGarrie and D.~C. Thompson, {\it {Warped General Gauge Mediation}},
  \href{http://xxx.lanl.gov/abs/1009.4696}{{\tt arXiv:1009.4696}}.

\bibitem{Drees:2004jm}
M.~Drees, R.~Godbole, and P.~Roy, {\it {Theory and phenomenology of sparticles:
  An account of four-dimensional N=1 supersymmetry in high energy physics}}, .
  Hackensack, USA: World Scientific (2004) 555 p.

\bibitem{Martin:1993zk}
S.~P. Martin and M.~T. Vaughn, {\it {Two loop renormalization group equations
  for soft supersymmetry breaking couplings}},  {\em Phys. Rev.} {\bf D50}
  (1994) 2282, [\href{http://xxx.lanl.gov/abs/hep-ph/9311340}{{\tt
  hep-ph/9311340}}].

\bibitem{Yamada:1994id}
Y.~Yamada, {\it {Two loop renormalization group equations for soft SUSY
  breaking scalar interactions: Supergraph method}},  {\em Phys. Rev.} {\bf
  D50} (1994) 3537--3545, [\href{http://xxx.lanl.gov/abs/hep-ph/9401241}{{\tt
  hep-ph/9401241}}].

\bibitem{Bhattacharyya:2010rm}
G.~Bhattacharyya and T.~S. Ray, {\it {A phenomenological study of 5d
  supersymmetry}},  {\em JHEP} {\bf 05} (2010) 040,
  [\href{http://xxx.lanl.gov/abs/1003.1276}{{\tt arXiv:1003.1276}}].

\bibitem{Schmaltz:2000gy}
M.~Schmaltz and W.~Skiba, {\it {Minimal gaugino mediation}},  {\em Phys. Rev.}
  {\bf D62} (2000) 095005, [\href{http://xxx.lanl.gov/abs/hep-ph/0001172}{{\tt
  hep-ph/0001172}}].

\bibitem{Schmaltz:2000ei}
M.~Schmaltz and W.~Skiba, {\it {The superpartner spectrum of gaugino
  mediation}},  {\em Phys. Rev.} {\bf D62} (2000) 095004,
  [\href{http://xxx.lanl.gov/abs/hep-ph/0004210}{{\tt hep-ph/0004210}}].

\bibitem{Auzzi:2010xc}
R.~Auzzi and A.~Giveon, {\it {Superpartner spectrum of minimal gaugino-gauge
  mediation}},  \href{http://xxx.lanl.gov/abs/1011.1664}{{\tt
  arXiv:1011.1664}}.

\bibitem{Auzzi:2010mb}
R.~Auzzi and A.~Giveon, {\it {The sparticle spectrum in Minimal gaugino-Gauge
  Mediation}},  {\em JHEP} {\bf 10} (2010) 088,
  [\href{http://xxx.lanl.gov/abs/1009.1714}{{\tt arXiv:1009.1714}}].

\bibitem{Green:2010ww}
D.~Green, A.~Katz, and Z.~Komargodski, {\it {Direct Gaugino Mediation}},
  \href{http://xxx.lanl.gov/abs/1008.2215}{{\tt arXiv:1008.2215}}.

\bibitem{Komargodski:2010mc}
Z.~Komargodski, {\it {Vector Mesons and an Interpretation of Seiberg Duality}},
   {\em JHEP} {\bf 02} (2011) 019,
  [\href{http://xxx.lanl.gov/abs/1010.4105}{{\tt arXiv:1010.4105}}].

\bibitem{Abel:2010vba}
S.~Abel {\em et.~al.}, {\it {Pure General Gauge Mediation for Early LHC
  Searches}},  {\em JHEP} {\bf 12} (2010) 049,
  [\href{http://xxx.lanl.gov/abs/1009.1164}{{\tt arXiv:1009.1164}}].

\bibitem{Abel:2009ve}
S.~Abel, M.~J. Dolan, J.~Jaeckel, and V.~V. Khoze, {\it {Phenomenology of Pure
  General Gauge Mediation}},  {\em JHEP} {\bf 12} (2009) 001,
  [\href{http://xxx.lanl.gov/abs/0910.2674}{{\tt arXiv:0910.2674}}].

\bibitem{Rajaraman:2009ga}
A.~Rajaraman, Y.~Shirman, J.~Smidt, and F.~Yu, {\it {Parameter Space of General
  Gauge Mediation}},  {\em Phys. Lett.} {\bf B678} (2009) 367--372,
  [\href{http://xxx.lanl.gov/abs/0903.0668}{{\tt arXiv:0903.0668}}].

\bibitem{Thalapillil:2010ek}
A.~M. Thalapillil, {\it {Low-energy Observables and General Gauge Mediation in
  the MSSM and NMSSM}},  \href{http://xxx.lanl.gov/abs/1012.4829}{{\tt
  arXiv:1012.4829}}.

\bibitem{Hebecker:2001ke}
A.~Hebecker, {\it {5D super Yang-Mills theory in 4-D superspace, superfield
  brane operators, and applications to orbifold GUTs}},  {\em Nucl. Phys.} {\bf
  B632} (2002) 101--113, [\href{http://xxx.lanl.gov/abs/hep-ph/0112230}{{\tt
  hep-ph/0112230}}].

\bibitem{Komargodski:2008ax}
Z.~Komargodski and N.~Seiberg, {\it {mu and General Gauge Mediation}},  {\em
  JHEP} {\bf 03} (2009) 072, [\href{http://xxx.lanl.gov/abs/0812.3900}{{\tt
  arXiv:0812.3900}}].

\bibitem{Buican:2008ws}
M.~Buican, P.~Meade, N.~Seiberg, and D.~Shih, {\it {Exploring General Gauge
  Mediation}},  {\em JHEP} {\bf 03} (2009) 016,
  [\href{http://xxx.lanl.gov/abs/0812.3668}{{\tt arXiv:0812.3668}}].

\bibitem{Benakli:2008pg}
K.~Benakli and M.~D. Goodsell, {\it {Dirac Gauginos in General Gauge
  Mediation}},  {\em Nucl. Phys.} {\bf B816} (2009) 185--203,
  [\href{http://xxx.lanl.gov/abs/0811.4409}{{\tt arXiv:0811.4409}}].

\bibitem{Intriligator:2008fr}
K.~A. Intriligator and M.~Sudano, {\it {Comments on General Gauge Mediation}},
  {\em JHEP} {\bf 11} (2008) 008,
  [\href{http://xxx.lanl.gov/abs/0807.3942}{{\tt arXiv:0807.3942}}].

\bibitem{Distler:2008bt}
J.~Distler and D.~Robbins, {\it {General F-Term Gauge Mediation}},
  \href{http://xxx.lanl.gov/abs/0807.2006}{{\tt arXiv:0807.2006}}.

\bibitem{Ooguri:2008ez}
H.~Ooguri, Y.~Ookouchi, C.-S. Park, and J.~Song, {\it {Current Correlators for
  General Gauge Mediation}},  {\em Nucl. Phys.} {\bf B808} (2009) 121--136,
  [\href{http://xxx.lanl.gov/abs/0806.4733}{{\tt arXiv:0806.4733}}].

\bibitem{Lee:2010kb}
J.~Y. Lee, {\it {Renormalization in General Gauge Mediation}},  {\em JHEP} {\bf
  10} (2010) 041, [\href{http://xxx.lanl.gov/abs/1001.1940}{{\tt
  arXiv:1001.1940}}].

\bibitem{Argurio:2010fn}
R.~Argurio, M.~Bertolini, G.~Ferretti, and A.~Mariotti, {\it {Unscreening the
  Gaugino Mass with Chiral Messengers}},  {\em JHEP} {\bf 12} (2010) 064,
  [\href{http://xxx.lanl.gov/abs/1006.5465}{{\tt arXiv:1006.5465}}].

\bibitem{Kang:2010ye}
Z.~Kang, T.~Liu, T.~Li, and J.~M. Yang, {\it {Semi-direct Gauge-Yukawa
  Mediation}},  \href{http://xxx.lanl.gov/abs/1012.4533}{{\tt
  arXiv:1012.4533}}.

\bibitem{Argurio:2009ge}
R.~Argurio, M.~Bertolini, G.~Ferretti, and A.~Mariotti, {\it {Patterns of Soft
  Masses from General Semi-Direct Gauge Mediation}},  {\em JHEP} {\bf 03}
  (2010) 008, [\href{http://xxx.lanl.gov/abs/0912.0743}{{\tt
  arXiv:0912.0743}}].

\bibitem{Csaki:2001em}
C.~Csaki, J.~Erlich, C.~Grojean, and G.~D. Kribs, {\it {4D constructions of
  supersymmetric extra dimensions and gaugino mediation}},  {\em Phys. Rev.}
  {\bf D65} (2002) 015003, [\href{http://xxx.lanl.gov/abs/hep-ph/0106044}{{\tt
  hep-ph/0106044}}].

\bibitem{Sudano:2010vt}
M.~Sudano, {\it {General Gaugino Mediation}},
  \href{http://xxx.lanl.gov/abs/1009.2086}{{\tt arXiv:1009.2086}}.

\bibitem{Martin:1996zb}
S.~P. Martin, {\it {Generalized messengers of supersymmetry breaking and the
  sparticle mass spectrum}},  {\em Phys. Rev.} {\bf D55} (1997) 3177--3187,
  [\href{http://xxx.lanl.gov/abs/hep-ph/9608224}{{\tt hep-ph/9608224}}].

\bibitem{Marques:2009yu}
D.~Marques, {\it {Generalized messenger sector for gauge mediation of
  supersymmetry breaking and the soft spectrum}},  {\em JHEP} {\bf 03} (2009)
  038, [\href{http://xxx.lanl.gov/abs/0901.1326}{{\tt arXiv:0901.1326}}].

\bibitem{Ghinculov:1994sd}
A.~Ghinculov and J.~J. van~der Bij, {\it {Massive two loop diagrams: The Higgs
  propagator}},  {\em Nucl. Phys.} {\bf B436} (1995) 30--48,
  [\href{http://xxx.lanl.gov/abs/hep-ph/9405418}{{\tt hep-ph/9405418}}].

\bibitem{vanderBij:1983bw}
J.~van~der Bij and M.~J.~G. Veltman, {\it {Two Loop Large Higgs Mass Correction
  to the rho Parameter}},  {\em Nucl. Phys.} {\bf B231} (1984) 205.

\bibitem{Intriligator:2010be}
K.~Intriligator and M.~Sudano, {\it {General Gauge Mediation with Gauge
  Messengers}},  {\em JHEP} {\bf 06} (2010) 047,
  [\href{http://xxx.lanl.gov/abs/1001.5443}{{\tt arXiv:1001.5443}}].

\bibitem{Matos:2010ie}
L.~Matos, {\it {Gauge Mediation with Gauge Messengers in SU(5)}},  {\em JHEP}
  {\bf 12} (2010) 042, [\href{http://xxx.lanl.gov/abs/1007.3616}{{\tt
  arXiv:1007.3616}}].

\bibitem{Chacko:1999mi}
Z.~Chacko, M.~A. Luty, A.~E. Nelson, and E.~Ponton, {\it {Gaugino mediated
  supersymmetry breaking}},  {\em JHEP} {\bf 01} (2000) 003,
  [\href{http://xxx.lanl.gov/abs/hep-ph/9911323}{{\tt hep-ph/9911323}}].

\bibitem{Kaplan:1999ac}
D.~E. Kaplan, G.~D. Kribs, and M.~Schmaltz, {\it {Supersymmetry breaking
  through transparent extra dimensions}},  {\em Phys. Rev.} {\bf D62} (2000)
  035010, [\href{http://xxx.lanl.gov/abs/hep-ph/9911293}{{\tt
  hep-ph/9911293}}].

\bibitem{Abel:2010vb}
S.~Abel and T.~Gherghetta, {\it {A slice of $AdS_5$ as the large N limit of
  Seiberg duality}},  {\em JHEP} {\bf 12} (2010) 091,
  [\href{http://xxx.lanl.gov/abs/1010.5655}{{\tt arXiv:1010.5655}}].

\bibitem{Bouchart:2011va}
C.~Bouchart, A.~Knochel, and G.~Moreau, {\it {Discriminating 4D supersymmetry
  from its 5D warped version}},  \href{http://xxx.lanl.gov/abs/1101.0634}{{\tt
  arXiv:1101.0634}}.

\end{thebibliography}

\providecommand{\href}[2]{#2}\begingroup\raggedright\endgroup

\end{document}